\documentclass[12pt,a4paper,oneside]{article}
\usepackage[margin=1in]{geometry} 

\usepackage{cite}
\usepackage{amsmath,amssymb,amsfonts}
\usepackage{algorithmic}
\usepackage{graphicx}
\usepackage{textcomp}
\usepackage{tabularx,booktabs}
\usepackage{float}      
\usepackage{caption}   
\usepackage{physics}
\usepackage{hyperref}
\usepackage[nobiblatex]{xurl}
\PassOptionsToPackage{hyphens}{url}

\usepackage{authblk}

\usepackage{lscape} 

\title{Bridging the Gap to Next Generation Power System Planning and Operation with Quantum Computation}

\author[1]{Priyanka Arkalgud Ganeshamurthy\thanks{e-mail: priyanka.ag@eonerc.rwth-aachen.de}}
\author[2]{Kumar Ghosh}
\author[2]{Corey O'Meara}
\author[2]{Giorgio Cortiana}
\author[3]{Jan Schiefelbein-Lach}
\affil[1]{E.ON Energy Research Center, Institute for Automation of Complex Power Systems, RWTH Aachen University, Aachen, Germany}
\affil[2]{E.ON Digital Technology GmbH, Hannover, Germany}
\affil[3]{E.ON Group Innovation GmbH, Essen, Germany}
 
\begin{document}


\maketitle




\begin{abstract}
Innovative solutions and developments are being inspected to tackle rising electrical power demand to be supplied by clean forms of energy. The integration of renewable energy generations, varying nature loads, importance of active role of distribution system and consumer participation in grid operation has changed the landscape of classical power grids. Implementation of smarter applications to plan, monitor, operate the grid safely are deemed paramount for efficient, secure and reliable functioning of grid. Although sophisticated computations to process gigantic volume of data to produce useful information in a time critical manner is the paradigm of future grid operations, it brings along the burden of computational complexity. Advancements in quantum technologies holds promising solution for dealing with demanding computational complexity of power system related applications. 
In this article, we lay out clear motivations for seeking quantum solutions for solving computational burden challenges associated with power system applications. Next we present an overview of quantum solutions for various power system related applications available in current literature and suggest future topics for research. We further highlight challenges with existing quantum solutions for exploiting full quantum capabilities. Additionally, this paper serves as a bridge for power engineers to the quantum world by outlining essential quantum computation fundamentals for enabling smoother transition to future of power system computations.  
\end{abstract}



\newpage
\tableofcontents

\newpage
\label{sec:introduction}
\section{Introduction}


Electrification of sectors such as mobility and heating has led to an increase in demand on today's power grid. In order to support this growing demand, power sector is taking many steps such as inclusion of additional generation resources, expansion of grid infrastructure, strengthening infrastructure of information and communication technology (ICT) , innovation in smarter grid operational solutions and exchange of immense volume of data for efficient and safe operation of the grid. Handling and processing enormous volume of data using sophisticated algorithms within the speed which is dictated by the specific application can be computationally expensive on today's processors.   
In recent years there has been increasing efforts from both industries and research institutions to explore Quantum solutions for power system applications to address this challenge \cite{QC_PA_01}, \cite{QC_PA_10}, \cite{QC_PA_06}. To this end, this paper aims to serve as a primer for bridging the knowledge gap between power system engineers and quantum specialists by adequately highlighting the need for research in this direction followed by providing comprehensive outlook on existing quantum solutions and algorithms for applications used in planning and operation of power systems as well as suggesting it's quantum readiness level.  
The main contributions of this paper are:

\begin{itemize}
    \item  to outline clear motivations for continuing research on quantum solutions for power system applications,
    \item to present an overview of current research on quantum-solutions for different power system, applications and suggest quantum readiness level
    \item to outline challenges in exploiting full quantum potential specifically for power system applications, 
    \item to present quantum fundamentals essential for developing a quantum algorithms,
    \item to identify potential applications which can benefit from quantum computations 
\end{itemize} 


Intended audience of the paper are (but not restricted to) power system engineers and quantum specialists. For power system engineers this paper serves three purposes. Firstly, to drive attention of power engineers and motivate for investigating quantum solutions to address emerging computational challenges of various complex power system applications. Second, to present quantum approaches investigated for power system related applications in current literature. Third, to provide fundamental concepts of quantum computation and right references for further information, in an effort to build the gap of quantum knowledge. On the other hand, quantum speacialists may benefit to understand the technical and computational challenges associated with various power system applications and thereby demanding their attention to investigate power system specific quantum solutions.

\section{Why Quantum Compution for Future Power System Applications}

To address the increasing need for decarbonization for environment sustainability, the energy sector is moving at a fast pace towards the use of decentralized and renewable energy resource for electricity production \cite{RES_news_01}. Further, with increase in energy demand, integration of new types of loads coupled with variability and intermittent nature of generation and loads is transforming not only the dynamics of power system, but also increasing the complexity of applications required to support the modern power network. 

Increased integration of renewable based generation has also motivated phasing out of conventional fossil-fuel based generation. The conventional generators such as synchronous machines have rotating masses that introduce sufficient inertia to the system, which is crucial for the stability of the power system. On contrary, the renewable sources are integrated to the grid through power electronics based converters, which does not introduce inertia to the system. Therefore, in grids with high proportions of renewable resources integration, the grid experiences faster rate dynamics due to low-inertia of power system. This demands faster execution of power system applications to support real-time grid monitoring, operation and control. Moreover, the volume of data needed for effective monitoring and optimized operation of modern grid has increased tremendously. Therefore, today's processors for power system applications are faced with the challenge to process huge amount of data and process complex computation in close to real-time. 

The main computational challenge for most power system applications stem from:
\begin{itemize}
    \item increase in number of variables due to expanding grid size \footnote{Although the operation and control of the grid takes place region wise, when the medium and low-voltage section of the grid is to be included, then the number of variables to be computed and number of resources to be considered for the monitoring, operation, control and optimization problem increases.} and grid components,
    \item complexity in executing algorithms corresponding to mathematical formulation of problems,
    \item handling volume of data needed to perform the computation and generated by the computation,
    \item frequency with which computation needs to be performed,
    \item and speed with which computation outcomes are desired.
\end{itemize}

Table \ref{lit_rev_01} provides an indication the source of computational complexity for different power system application categories. 

\begin{table}[!htbp]
\centering
\caption{Indication on power system applications computation complexity}
\label{lit_rev_01}
\scalebox{1}{
\begin{tabular}{l l c }
\toprule

\multicolumn{1}{c}{Power System } & \multicolumn{1}{l}{Sub-Category} &  \multicolumn{1}{c}{Computation}   \\

\multicolumn{1}{c}{Application} & \multicolumn{1}{l}{} &  \multicolumn{1}{c}{Complexity} \\

\cline{1-3}
 Grid Situational& Static States Estimation & MR \\
Awareness& Dynamic States Estimation & MR, RT \\
& Parameter estimation & MR  \\
 & Observability Analysis & MR  \\
 & Meter Placement & {O} \\
 & Power flow & - \\

\cline{1-3}
Grid Security & Contingency Analysis & MR \\
 & Cyber Security & - \\
 & Reliability Analysis & MR \\
\cline{1-3}
Optimization & Unit commitment & O, MR \\
& Economic power dispatch & O, MR \\
& Facility location allocation & O \\
& Volt/Var control & O, MR \\
\cline{1-3}
 Forecasting & Weather forecast & MR \\
& Generation forecast & MR \\
& Load forecast & MR \\
& Electricity price forecast & MR, RT \\
\cline{1-3}
Grid Stability & Small signal Analysis & MR \\
 & Transient Analysis & MR \\
 & EMTP & MR \\
 & Fault Detection & RT  \\
\cline{1-3}
Grid Control & Load frequency control & O, MR, RT \\
& DFIG rotor control & O, MR, RT\\
\bottomrule
\end{tabular}
}
  \begin{tabular}{ll}   

    \small{MR} & Multiple execution runs           \\ 
    \small{RT} & Faster or close to real-time execution desired \\
    \small{O} & Combinatorial and/or constrained optimization\\
       
  \end{tabular}

\end{table}

Execution of computationally burdened  power system applications on classical computers is often faced with the trade-off between execution speed and outcome accuracy. Generally in order to gain higher execution speed, the problem formulations are adapted with certain approximations in order to simplify the complexity of the problem, which leads to lower accuracy. One obvious solution could be developing classical computers with much higher processing power, such as Tianhe-2 supercomputer. However, these supercomputers demands extraordinary level of energy consumption \cite{QC_PA_10}, \cite{QC_news_03}.  

\begin{figure}
    \centering
    \includegraphics[width=0.75\textwidth]{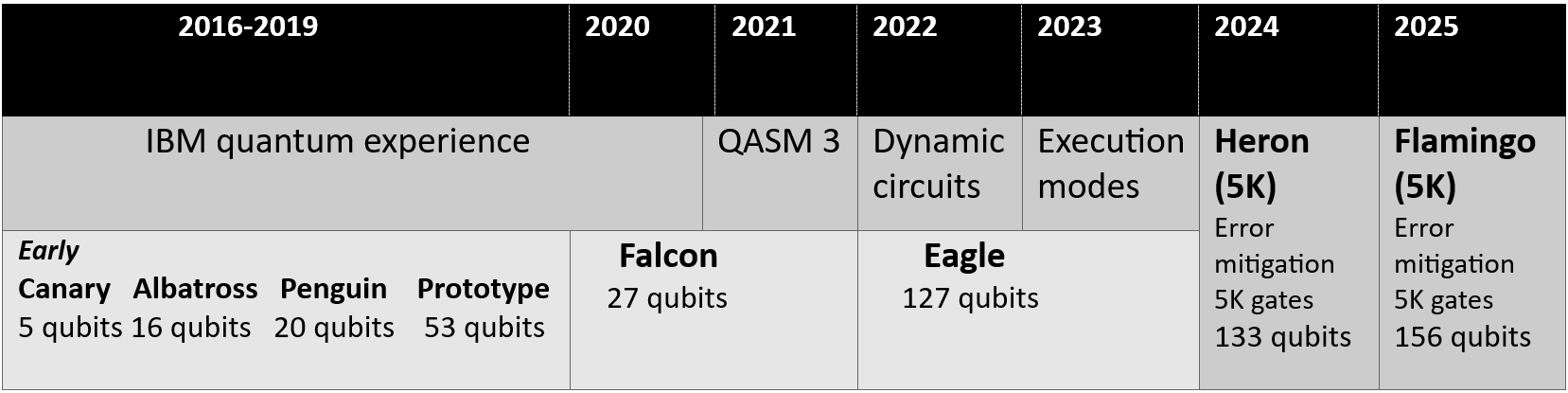}
    \caption[Caption used in list of tables]{IBM Quantum development roadmap \cite{QC_news_02}.}
    \label{QC_RES}
\end{figure}

An alternative possible solution to this lies in another important technology which has gained traction in recent years, that is, the Quantum computing. The increase in number of Qubits, reduction in noise and formulation of critical algorithms are paving way for improved execution of existing engineering applications with Quantum computation. As an example, Fig. \ref{QC_RES} shows development roadmap of IBM Quantum technology \cite{QC_news_02} over the years. Quantum computing is capable of administering large data sets at much faster speeds, thereby rising as a promising solution to the problem of computational burden of power system applications.   
Moreover, quantum computers promises use of significantly less power than a classical computer as speculated in \cite{QC_news_04}, \cite{QC_news_01}. 


While Quantum computation is a promising solution, encoding of information into a Quantum computer and processing of quantum information is very different in comparison to its classical counterpart. Quantum computing adoption therefore poses two-fold challenge. First to train power sector engineers with specialized skill for quantum computation with a clear understanding of the technology’s capabilities and limitations. Second to build quantum algorithms and proof-of-concepts addressing various power system applications. This will not only ensure a smoother transition to the future grid system, but also enable the power community to be future-ready. 

With this background, the main motivation for exploring the quantum solutions for power system applications are the following:
\begin{itemize}
    \item Capability to handle and process huge volume of data
    \item Capability for real-time execution of required applications
    \item Possible solution for gaining computation speed without compromising on accuracy 
    \item Decrease computation-driven carbon footprint
    \item Develop new algorithms addressing power system applications, enabling future-preparedness
    \item Train power engineers with appropriate Quantum computation related skill sets, enabling smoother transition
\end{itemize}

\section{Application of Quantum Computing for Power Grids}
In this section, we review existing Quantum proof-of-concept implementations of crucial functions for power system management and safe operation.


\subsection{Grid Situational Awareness}

Situational awareness of the grid is extremely important for the grid operators in order to be able to operate the grid in reliable and safe manner. The state of the grid is monitored using the live measurement data from the field and historical data on the load and generations. The grid state monitoring using weighted least squares technique is a common approach followed in the transmission grid operations. In recent years, distribution grid monitoring is equally gaining importance due to the transformation of distribution grid from once passive to now active grid. 
In this sub-section, we review few sub-routines that are used to support the monitoring of power system and their corresponding quantum-enabled approaches proposed in literature:

\subsubsection{Static states estimation}
 State estimation routine is one of the most critical routine in the control center, as its results are further utilized in other energy management system routines such as optimal power flow, contingency analysis to name a few. 

States estimation involves the problem of estimating the quantities of the grid, with whose knowledge the complete state of the grid can be known. The most common selection of states is the voltage magnitude and phase angles. For a grid with $n$ buses, $2n-1$ states needs to be computed. Therefore, the problem scales in complexity with size of the grid. 
 
 Static state estimation is a category of state estimation, where the state of system at time instant $t$ is estimated using measurements at time $t$, and does not depend on the state of the grid at $t-1$. One of the most popular formulation of static state estimation problem is based on Weighted Least Squares (WLS) method, where the objective is to minimize square of the measurement residuals, as given in \eqref{wls}. Owing to the non-linear nature of the problem formulation, the method involves Gauss-newton iterations to compute the states, as shown in \eqref{wls_gauss_iter}. 
 
\begin{align}
g(x) = -H^T(x)W[z-h(x)] = 0
\label{wls}
\end{align}

\begin{align}
    \begin{split}
    g(x_{k+1}) &= g(x_k) + G(x_k)(x_{k+1} - x_k) = 0\\
    G(x_k).\Delta{x} &=  -H^T(x_k)W[z-h(x_k)]
    \end{split}
    \label{wls_gauss_iter}
\end{align}

 Computationally intensive steps in WLS method include calculation of Jacobian matrix $H(x)$ and the inversion of the Gain matrix $G(x) = H(x)^TWH(x)$ in each iteration of the method. In case of transmission grid, the fast decoupled method can be used which involves making certain assumptions, which avoids calculation of Jacobian and inversion of Gain matrix in each iteration, and therefore reducing the computational burden. However, such assumptions do not hold true in case of the distribution grids, and therefore the problem remains computationally challenging. Further adding to the complexity of the problem is the unbalanced nature of distribution grid, which calls for estimation of states for all three-phases. In addition, the distribution grids contains a larger number of buses in comparison to the transmission grid. These challenges today's available computers capability to compute states estimation for growing distribution grids of future. 
Recent research in \cite{QC_SA_02} investigates Quantum algorithms to solve WLS method. The method proposes Harrow-Hassidim-Lloyd (HHL)-enabled Quantum implementation of WLS for  microgrids. Further, to deal with ill-conditioned gain matrix, Authors of \cite{QC_SA_02} present preconditioned quantum linear solver. The presented methods provide a firm basis and reveals the potential of quantum computing in microgrid state estimation, however due to the fact that current quantum computers still have limitations regarding the quantum depth, coherence time, and noise tolerance capability, the presented investigation is limited to a small size of microgrid. Quantum assisted AC three-phase distribution grid state estimation, and hybrid AC-DC states estimation may be considered for future investigations. 

\subsubsection{Dynamic states estimation}
\label{sec: DSE}
Dynamic state estimation (DSE) tracks the dynamics of power system states and provides evolution of the system state with time. Computation of states using dynamic state estimation method at time $t$ depends not only on the measurements at $t$, but also on the states at time $t-1$ and therefore gives a picture of evolution of state of the grid. Widely used classes of methods for dynamic state estimation is based on Kalman Filtering techniques, including extended kalamn filter and unscented kalman filters. A kalman filter uses a series of measurements over time to produce estimates of unknown variables based on a dynamic model. Recent IEEE Task force efforts \cite{C_SA_02}, \cite{taskforce4} has highlighted the importance of dynamic states estimation for the emerging grid conditions and for applications ranging from monitoring, operation, control, protection and planning. The majority of DSE designs recursively compute gains $K$, as given by \eqref{dse}; this involves matrix multiplications and inversions at each time-step. DSE need to be supported by adequate computing resources to achieve practical and acceptable performance in real-time \cite{C_SA_03}. 

\begin{align}
K = \Tilde{P}H^T(H\Tilde{P}H^T + R)^{-1}
\label{dse}
\end{align}

The use of Quantum computing for dynamic state estimation raises naturally as an attractive solution. For a quantum system, the stochastic filter is provided by a quantum filter, which is also known as a stochastic master equation (SME). For a linear quantum system subject to linear measurements and Gaussian noise, the quantum filter reduces to a quantum Kalman filter (QKF) \cite{QC_SA_05}. In this direction, \cite{QC_SA_04} has presented a Quantum algorithm to perform the extended kalman filtering, where a commutative approximation and a time-varying linearization to non-commutative quantum stochastic differential equations (QSDEs). Therefore using quantum extended kalman filtering (QEKF) for power system dynamic states estimation can be explored for future research.

\subsubsection{Observability Analysis}
\label{sec: Observability}
The aim of observability analysis is to check whether enough measurements are available to perform states estimation calculation and in case of unobservability, to find which section of grid states can be estimated with available measurements. The observability analysis can be performed offline, for planning of meter placement while expansion of grid, or performed online during operation, because of changing topology or loss of communication network can lead to unobservable grid. This may necessitate execution of observability analysis multiple times during the operation. 
    
   For static state estimation, the outcome of observability analysis is binary and there are broadly two categories of methods to perform this analysis, they are the topological observability analysis and numerical observability analysis \cite{Abur_Exposito}. The topographical observability analysis is based on representing the grid in a graph-based formulation, with buses representing graph nodes and lines represented by the graph edges. One of the algorithm to find observability using this graph-based representation is called Nucera's algorithm and does not involve any mathematical computation. Nucera's algorithm follows a set of assumptions and specific rules to find the whether the grid is observable and/or to find observable islands of the grid. The challenge with this method is that with increasing size of the grid, the problem becomes increasingly complex and not intuitive. Quantum computing algorithms are considered for several problems in graph theory.  \cite{QC_SA_01} lists Quantum algorithms to perform graph-based algorithms.  Therefore, Quantum-enabled algorithm for topological observability problem may be a possible research direction for further investigation. The numerical observability analysis involves observing the non-singularity of the Gain matrix and whether Jacobian has a full-rank, satisfying which indicates the full observability. To identify the unobserved islands, the numerical method makes use of solution to linearized WLS equation. 
   
   For dynamic state estimation, the grid observability results in outcome indicating whether the grid is strongly or weakly observable. Due to the non-linearity and time-varying nature of the problem, the methods based on lie-derivative \cite{taskforce4} needs to be used in every time step of the dynamic estimation, which can be computationally demanding. Here, the Author's foresees the possibility of using quantum algorithms such as HHL or VQLS the numerical observability analysis as a direction for further investigation. 

\subsubsection{Meter Placement Problem}
The meter placement problem aims at two challenges, first to improve the grid visibility and second to reduce the estimation errors. However, due to the large number of nodes in distribution networks, it is not economically possible to install meters in all nodes, and therefore necessitates development of routines to investigate optimal meter placement to overcome specific challenges \cite{C_SA_01}. This is generally an offline procedure during planning stages to assess optimal location for placement of measuring device in order to obtain complete visibility of the grid and/or improve the accuracy of the state estimation. In \cite{QC_SA_03}, through Quantum optimization technique an assessment is performed on the computational viability of the current generation D-Wave Systems 2000Q Quantum Annealer (QA) for this power systems design problem. The analysis suggested that the QC might outperform well-developed conventional optimization methods in this application; however, limitations in hardware and noise for large and densely linked graphs need to be addressed first.

\subsection{Grid Security Assessment}
Evaluation of power system security is necessary in order to develop ways to maintain system operation to ensure continuation of supply of electricity. Power system security is the ability of power system to operate successfully in an unexpected event such as failure or loss of a component. Security is generally evaluated by considering the probable components failure based on reliability analysis and analyzing the effect of loss of such components on the operating conditions of the grid via contingency analysis. Further, modern day power system is heavily reliable and interdependent with the information and communication technology (ICT) network, owing to its important role in transitioning traditional grid to smart grid. As much as the ICT network empowers digitization and smart operation of the grid, it also exposes our grid to the threats which are cyber in nature. A cyber-attack to the ICT infrastructure of the grid can cascade and propagate to the process operation level, and has the potential to interfere with the operation of the grid. Therefore, in today's grid, the power system security assessment should involve assessment on reliability of physical equipment and device, effect on operation of the grid in the event of loss of an power equipment and also security of the ICT network.

\subsubsection{Contingency Analysis}
A contingency is the failure or loss of an element (e.g. generator, transformer, transmission line.), or a change of state of a device (e.g. the unplanned opening of a circuit breaker in a substation) in the power system \cite{C_SeA_01}. Therefore contingency analysis is an application that uses a computer simulation to evaluate the effects of removing individual elements from a power system. Current electric utility operating policies (such as NERC's) require that each utility's power system must be able to withstand and recover from any "first contingency" or any single failure at every operational state of the grid \cite{C_SeA_01}. However, in future the policies may require the utilities to test and withstand for the second contingency, which would then lead to a higher number of contingencies to be tested for a given size of grid. Contingency analysis is run also online. Whenever the state of the grid changes, the contingency analysis is run to check for security against first contingencies. As such one may anticipate, the contingency analysis problem scales with the size of the grid and the level up to which the contingency assessment is to be performed. Basically, the contingency analysis runs potentially thousands of power flow, each considering a different contingency, and assessing the system's condition due to each contingency. Recently, Quantum computing solution for DC and fast decoupled method of AC power flow solution has been investigated in \cite{ QC_SeA_04, QC_SeA_04_1} and \cite{QC_SeA_05}. The approaches make use of HHL algorithm for solving system of linear equations to solve the power flow problem. \cite{QC_SeA_02} introduces Quantum algorithm for the contingency analysis problem, where the HHL method is reformulated such that the method can be applicable for AC power flow problems. In \cite{QC_SeA_15}, authors present experimental results of AC power flow using HHL algorithm on real noisy intermediate-scale quantum computer (NISQ). The paper examine the impact of current noisy quantum hardware on the accuracy and speed of an AC power flow algorithm. The paper concludes that the current hardware is capable of performing a power flow for small test systems, but scalability is currently a major issue which needs further improvement.

\subsubsection{Cyber Security} 
Cyber security in power grid is an extensive and important area of research owing to the capability of cyber attacks to disrupt operation of electrical grid and thereby impacting nation's security and economy. While Quantum computing presents promising advantages in terms of computational efficiency and speed, these strengths pose as a challenge to the security of grid communication network. The security of public-key cryptographic systems heavily relies on the computational difficulty of specific mathematical problems such as  discrete logarithm and factoring problems \cite{QC_SeA_07}. However, in the age of improving Quantum technology, as reported in \cite{QC_SeA_06}, the Quantum-enabled attacks unleashes exploitation of new security vulnerabilities. For instance, the security of RSA (Rivest–Shamir–Adleman) is threatened by the Shor algorithm that can efficiently calculate prime factorization using Quantum computing, and on the other hand for symmetric encryptions, the brute-force attacks is made possible with increasing computational speed of quantum computers. This poses a challenge to the security of data exchanged over communication network amongst various stakeholders in the smart grid involving integration of Distributed Energy Resources (DER). In this regard, \cite{QC_SeA_06} presents investigation on defense strategies using quantum key distribution (QKD) and post-quantum cryptography (PQC). In \cite{QC_SeA_14}, QKD is used to authenticate smart grid communications, and is further demonstrated on electric utility's fiber network. Studies in \cite{QC_SeA_08}, \cite{QC_SeA_09}, \cite{QC_SeA_10}, \cite{QC_SeA_11} investigates application of QKD for enabling security in microgrids and networked microgrids. To overcome challenges, such as point-to-point and  distance limitation issue \cite{QC_PA_02} which hinder wide implementation of QKD for power grids, authors of \cite{QC_SeA_12} present Quantum networks as a possible solution. In addition to QKD, quantum direct communication (QDC) is another form of quantum communication, where confidential information is directly transmitted over Quantum channel. \cite{QC_SeA_13} utilizes a quantum-direct communication (QDC) approach to provide very high secured communication for resiliency improvement of power grid. 
To address the cyber-security challenges in case of microgrid distributed control, \cite{QC_DC_01} and \cite{QC_DC_02} presents a quantum distributed control framework to enable controlling networks of DERs through a network of quantum systems.

\subsubsection{Reliability Assessment} 
The reliability of a system is the probability with which the system will perform its required function under given conditions for a specified period of time. The reliability assessment makes use of failure rates and outage data of various components in the grid to assess the overall reliability of the power grid. Monte-carlo (MC) method is used as a simulation method for reliability assessment, where failure probability of elements of the system is estimated using the probability distribution histogram from Monte Carlo simulation. In MC method, first, the uncertain parameters are modeled as random variables with a given uncertainty distribution, then random samples are generated for each random variable using respective probability distribution function, and finally the expected value for each random value is calculated. The accuracy of MC method mainly depends on the number and quality of sampling. For a large-scale system, the required number of sampling is very large, which makes the method computationally challenging. Recently in \cite{QC_SeA_01} and \cite{QC_SeA_03} Quantum-enabled methods for MC simulation for reliability assessment have been proposed. Quantum amplitude estimation (QAE) has been used in \cite{QC_SeA_01} to execute the quantum circuit containing the probability distribution loading block for random variables total-time to failure (TTF) and total-time to repair (TTR), quantum MC simulation (QMCS) block, estimation of required functions of random variables, and measuring units. The results indicate that the  quantum  MC simulation algorithm needed a small number of qubits to produce the same results as compared to the classical MC simulation method in which the sampling size is an important factor in the convergence rate of the method.

\subsection{Optimization for Grid Application}
The planning and scheduling tasks in power system are generally formulated as an optimization problem, whose objectives can range from loss minimization, to generation cost minimization or maximization of revenue from electricity generation. The traditional planning and scheduling optimization problem are combinatorial in nature and are often NP-hard. The use of quantum optimization is expected to achieve a super-polynomial advantage for complicated combinatorial optimization problems \cite{QC_PA_02}. Currently available quantum optimization algorithms can already solve problems formulated as quadratic unconstrained binary optimization (QUBO) problems. One of the first quantum optimization algorithms is the Quantum Approximate Optimization Algorithm (QAOA). 

In literature, Quantum-enabled formulations of unit commitment (UC) problem is widely investigated \cite{QC_O_01}, \cite{QC_O_02}, \cite{QC_O_03}, \cite{QC_O_04}, \cite{QC_O_05}. The unit commitment is used to determine the start-up and shut-down schedule of all production units so that the electric demand is supplied and the total operating cost is minimized, at the same time meeting a number of system and generator constraints. The UC problem is generally formulated as a large-scale mixed integer nonlinear problem and involves nonlinear cost function and the feasible solutions are combinatorial in nature \cite{QC_O_01}. In studies \cite{QC_O_02}, \cite{QC_O_03}, \cite{QC_O_04}, \cite{QC_O_05}, different Quantum-enabled approaches to the unit commitment problem is formulated using QAOA. To fit the requirement of QAOA, \cite{QC_O_02}, \cite{QC_O_03} translate the UC model into sub-problems, where the binary  variables (commitment status of generators) are formulated by QUBO sub-problems, and the sub-problems are coordinated using the alternating direction method of multipliers (ADMM) \cite{QC_PA_02}. In \cite{QC_O_01} the mixed-integer quadratic programming (MIQP) UC problem, is reformulated as a QUBO problem and solved using a D-wave quantum Annealer (QA). {Beside ADMM, there is another decomposition called Benders decomposition which is very effective for solving the mixed integer optimization problems, where the whole optimization problem is decomposed in to a Master problem which is entirely integer-problem and a set of sub problems which are continuous optimization problems. In the recent years classical-quantum hybrid Benders decomposition \cite{BENDERS, Rahmaniani} approach has been developed to solve generic Mixed-integer-Linear programming (MILP) \cite{Zhao} and also UC problems \cite{Gao} where the Integer programming part is solved with QAOA or D-wave and the continuous optimization part is solved with a standard classical solver. Apart from the decomposition approach  a novel method called Bayesian optimization (BO) which is very efficient for optimizing complicated black-box reward functions. Recently BO method  has been applied for data-driven unit commitment problems \cite{Nikolaidis}. As an extension of classical BO, a quantum version of Bayesian optimization method \cite{Rapp2023, Dai2024} using quantum kernels has been developed to solve  some real-world optimization problems.}

Once the commitment of each generating unit is decided by the UC, the next step is to determine, for each hour of the planning horizon, the actual power output of each of the committed generating units that is needed to supply the demand and to comply with the set operational equality and inequality constraints. This is the economic power dispatch problem, where  the total generation costs are optimized. To address this problem, \cite{QC_O_06} proposes Quantum-behaved particle swarm optimization (QPSO) with Cauchy distribution. The paper presents results of QPSO on 15 generation unit network and compares the outcomes with QPSO with Cauchy distribution. 
 
Some other notable applications with optimization problem in distribution grids are:

\begin{itemize}
    \item Volt/Var control - to find optimal settings of control variables, such as generator voltages, transformer taps and shunt VAR compensation devices for effective voltage and reactive power control. \cite{QC_O_07} and \cite{QC_DC_04} propose use of quantum genetic algorithm (QGA) to address this problem to solve for optimal control variable set points.  
    \item Facility location-allocation - to find optimal place for installation of generation facilities in order to minimize capital and operational costs.  \cite{QC_O_01} presents approaches using VQE on IBM Quantum system and QA on D-Wave 2000Q system for addressing this problem. 
    \item Grid partitioning - to optimally partition the grid for easier analysis and operation of larger networks. \cite{QC_O_08} proposes application of quantum annealing for complex network theory based graph partitioning using electrical modularity.
    \item Service restoration 
    \item Optimal power flow for minimizing losses
    \item Redispatch for congestion management
     \item Optimal meter placement problem
\end{itemize}

\subsection{Forecasting}
\label{sec: Forecast}
Forecasting methods are used extensively in the DER integrated modern grids, with applications ranging from load forecasting, DER generation forecasting, weather forecasting and energy price forecasting for market functions. Machine learning techniques are widely used for forecasting and prediction problems. Classically, machine learning involves significant computational burden and its performance heavily depends on the choice of learning models. Since quantum states can be efficiently operated in the Hilbert space and are capable of representing entangled correlations, Quantum Machine Learning is promisingly powerful for data processing and model  training in ultra-high dimensional space that are intractable for classical algorithms \cite{QC_PA_02}.

References \cite{QC_F_01}, \cite{QC_F_02} and \cite{QC_F_03} reports and investigates Quantum machine learning based forecasting applications for power system.  

\subsubsection{Weather, Generation and Load Forecasting}
The need for weather, generation and load forecasts are increasing for efficient and safe grid operations in renewable dominated power networks. In \cite{QC_F_01}, a quantum counterpart of the support vector algorithm (QSVA) is proposed that can be used to forecast solar irradiation, whereas \cite{QC_F_02} presents Quantum Generalized Neural Network (QGNN) method for forecasting of solar photovoltaic system power output, wherein QC was implemented with the genetic algorithm to optimize the algorithm during the training process. In \cite{QC_F_03} a wind power prediction method based on the combination of quantum genetic algorithm and fuzzy neural network is presented to predict the wind power of wind farm in the short term. 

For the application of electric load forecasting, \cite{QC_F_04} presents a hybrid electric load forecasting technique using a support vector regression (SVR) algorithm, in which the QC mechanism is integrated with SVR to improve the forecasting accuracy. In this paper, quantum computing mechanism is used to quantamize dragonfly behaviors to enhance the searching effectiveness of the dragonfly algorithm, namely (QDA). Similarly, numerous preliminary trials have been proposed, in which Quantum computing mechanism and meta-heuristic algorithms are hybridized with an SVR model, such as the quantum PSO (QPSO) algorithm with SVR \cite{QC_F_05}, \cite{QC_F_06}, the quantum bat algorithm (QBAT) with SVR \cite{QC_F_07}. 

\subsubsection{Electricity Pricing}
Quantum systems could also play a vital role to optimize energy procurement, trading, and hedging, to better anticipate prices and power demand. For the purpose of electricity price forecasting, \cite{QC_F_08} utilizes quantum immune optimization algorithm (QIOA) and a modified back propagation neural network (BPNN) price prediction method. The proposed algorithm is studied on a realistic New Zealand power company, where the numerical results suggest higher prediction accuracy of proposed quantum immune optimization BP algorithm, compared to the traditional BPNN.

\subsubsection{Predictive Maintenance}
Another prediction based application investigated for benefiting the advantages of quantum computation is the predictive maintenance of gas power plants in \cite{QC_F_09}. Predictive maintenance method determine the condition of in-service equipment and estimate when maintenance of the monitored equipment should be performed. In \cite{QC_F_09}, a hybrid classical-quantum Autoencoder (HAE) model that performs anomaly detection for predictive maintenance is presented. The HAE is composed of a classical encoder, a parameterized quantum circuit and a classical decoder. The results indicate higher precision of the proposed approach over its classical counterpart.

\subsection{Grid Stability Analysis} 
Power system stability is broadly defined as that property of a power system that enables it to remain in a state of operating equilibrium under normal operating conditions and to regain an acceptable state of equilibrium after being subjected to disturbance \cite{C_St_03}. In many cases, instability and eventual loss of synchronism are initiated by some disturbance in the system resulting in oscillatory behavior that, if not damped, may eventually build up \cite{C_St_02}. Further, with converter based resources integrated into the grid, the grid has lower inertia owing to less damped system and higher frequency oscillations. 

\subsubsection{Small Signal Stability Analysis}
\label{sec: small_signal}
Generally, the small signal and large signal stability analysis are performed to analyze ability of the system to maintain synchronism under small disturbance and large disturbances respectively. 
In small signal analysis, the ability of the system to maintain synchronism under small disturbances, such as load changes, are studied. The widely used method for small-signal stability analysis is known as the modal analysis. In this method, the system is linearized around an operating point and the eigen values of the system are computed and analyzed to understand the oscillatory modes available under small disturbances, which dictates the small signal stability characteristics of a system. To compute eigen values of a matrix on a quantum computer, \cite{QC_St_02} proposes algorithms by combining quantum linear solver (QLS), quantum singular value estimation (QSVE) and quantum phase estimation (QPE). The proposed quantum algorithms are applicable for diagonalizable normal matrices and diagonalizable matrices whose eigenvalues have non-positive imaginary parts. However, the applicability of this approach for specific case of power system modal-analysis is pending to be investigated.

\subsubsection{Large Signal Stability Analysis}
In case of transient stability analysis, comparatively severe disturbances, such as faults, are applied and time domain simulations are performed to analyze system's stability under these disturbances. The classical transient stability analysis is based on numerical integration methods for performing time-domain simulation, which can be time consuming with increase in grid size. For  simulation and analysis of electromagnetic, electromechanical transients, the Electromagnetic Transient Program (EMTP) is a widely used tool \cite{C_Sim_01}. At the core of the EMTP lies the problem complexity of numerical integration, which is solved by applying the trapezoidal discretization at each time step to transform the dynamic equations of a power network into  numerical equations of an equivalent resistance network. The computational burden of this problem scales polynomially with the size of the grid. To address this challenge, \cite{QC_Sim_01} proposes a Quantum-enabled EMTP algorithm, which essentially uses HHL algorithm to solve the Quantum Linear System Problem. Due to the complexity of Quantum circuits used in the HHL based algorithm of \cite{QC_Sim_01}, the method may not be executed correctly on presents day NISQ machine owing to noise, and therefore making this approach suitable for noise-free ideal quantum machines. Further \cite{QC_Sim_02} develops a VQLS-enabled QEMTP algorithm which is practical and noise-resilient approach for EMTP analysis for presently available NISQ devices \cite{QC_PA_02}.

Data-driven  methods provide an alternative path where offline-trained neural networks are used to establish stability regions \cite{C_St_01} and post disturbance conditions are compared to determine stability of the system post disturbance. This approach is used in \cite{QC_St_01}, where the transient stability features are embedded into quantum states through a variational quantum circuits (VQC) which serves as Quantum neural networks (QNN). The method is tested on both Quantum simulator and real IBM Quantum device.

\subsubsection{Fault Management}
In an event of a fault, the fault management functions such as fault detection, fault isolation and service restoration are critical in ensuring grid stability and continuity of supply. In this direction, \cite{QC_FM_01} presents Quantum computing-based deep learning framework for fault diagnosis, where the computational challenges due to the complexities of deep learning models are overcome by QC-based training.

\subsection{Grid Control}

Controllers such as proportional-integral (PI) and proportional-integral-derivative (PID) are used widely for various control applications. One such application is the load-frequency control to maintain power balance, for which secondary controllers are introduced to regulate the power system parameters within a specified limit during sudden load demand period. For better controlled response under load variations, a suitable controller gain value selection is considered essential. Several optimization techniques are implemented to optimize the secondary controller gain values. The existing techniques for controller gain selection would be significantly challenging in large and complex power systems due to non-linearity. To address this challenge, Quantum -enabled algorithms for application of tuning controller gains are investigated in the literature. In \cite{QC_DC_03} Genetic Algorithm (GA), Quantum Inspired Genetic Algorithm (QIGA) and Quantum Inspired Evolutionary Algorithm (QIEA) are proposed for tuning of controller gain values of a three area single stage reheat thermal power systems. 

For the purpose of rotor control in doubly-fed induction generator (DFIG), \cite{QC_DC_05} proposes a quantum parallel multi-layer Monte Carlo optimization algorithm (QPMMCOA) to achieve maximum power and improved generation efficiency by optimizing the PI controller parameters of the rotor-side converters. For similar application, however to avoid online optimization process and update control strategy online, \cite{QC_DC_06} proposes an an online control algorithm based on the quantum process, deep belief networks, and reinforcement learning. The proposed method can update the control strategy online with general initialization for dynamic systems, avoid optimal local solutions, and predict the next systemic states of DFIG. The Authors refer to the proposed approach as a quantum deep reinforcement learning (QDRL).

From this literature survey the authors have determined that the following power system applications lack a quantum proof-of-concept in the literature but have the potential to benefit from quantum computation.
\begin{itemize}
    \item Dynamic state estimation - As highlighted in Section \ref{sec: DSE}, DSE can be useful  in providing dynamic grid situational awareness to the operator for converter driven grids, however it suffers from computational burden due to the need for faster execution of costly computations, frequency of execution and size of the grid.  Therefore DSE can be foreseen to  benefit from quantum algorithms such as HHL, however a proof-of-concept for DSE for power system monitoring needs to be developed and tested for a reasonable grid size.  
    \item Dynamic observability analysis: The dynamic observability of the grid can vary at different times due to non-linearity and time variant nature of the power system. Due to this, the observability analysis needs to be executed frequently, using methods such as Lie derivatives. The computational load of these methods is affected by size of grid, number of measurements and the frequency of observability analysis execution. As highlighted in Section \ref{sec: Observability}, potential quantum algorithms can be investigated and tested for observability analysis. 
    \item Small signal stability analysis - The computational demand in modal analysis can arise from the size of the matrix and number of operating points for which the small signal stability needs to be tested. As mentioned in Section \ref{sec: small_signal}, modal analysis  can benefit from quantum methods to identify the eigenvalues of a matrix, however further research in this direction is required.   
    \item Service restoration - The goal of any service restoration algorithm is to ensure that all loads are supplied with power after an event of failure and isolation of faulty section. In doing so, the algorithm needs to solve a constrained and combinatorial optimization algorithm to arrive at an optimum way in which the electricity supply for non-faulty section is restored. Therefore, the computational burden of this algorithm is directly affected by the grid and component size, as well as by the requirement to be executed in a timely manner for ensuring minimum operational losses. The Authors foresee the use of quantum optimization algorithms in aiding speed in service restoration algorithms.  
    \item Re-dispatch for congestion management - Similar to service restoration algorithms, the congestion management algorithms have to solve a constrained optimization problem for ensuring optimum set-points for generating units and network configuration for reducing the congestion in grid. Here as well the execution of such algorithms face computational challenge due to the size of grid and components, which increase number of combinations.   
    \item Energy data analytics - 
    The field of big data, statistical software and machine learning (ML) techniques is gaining traction for use in energy data analytics for not only improving grid efficiency but also to improve service to customers. However these algorithms suffers from processing load due to huge volume of data and complexity in problem formulation. Apart from those presented in Section \ref{sec: Forecast}, several other applications such as demand response, load management, asset management and customer analytics can be foreseen to benefit from quantum computation advantages. 
\end{itemize}

\subsection{Challenges to Exploit Full Potential of Quantum Computation for Power System Applications}
Although quantum computers have a potential to outperform the classical computers for various computational problems, there are several challenges which limit exploitation of full quantum potential. 

\begin{itemize}
    \item Most of the popular quantum algorithms, e.g.,  Shor's \cite{Shor} or Grover's \cite{Grover} algorithms, require fault-tolerance, which is currently not available with state-of-the-art technology. 
    \item In the current \textit{noisy intermediate-scale (NISQ)} era, quantum computers have qubit-limitation for implementing the error correction protocols, making most of the quantum algorithms unsuitable for near-term devices.
    \item Validation of most of the quantum solutions for grid applications are limited to a small grid size due to limitations in quantum resources, since {a large quantum resources (in terms of qubits and circuit depth) are required for computation of bigger grids. The research so-far is limited to only 3-4 node systems. For example, a simple DC power flow problem executed using HHL, where the number of quantum gates grows enormously  as the system size grows, which is beyond the scope of current hardware resources.} 
    \item In the current status of quantum hardware, there are few limitations on quantum memory for efficiently encoding data. Quantum memory like qRAM is very hard to experimentally construct. The other approach is to directly prepare the quantum state through quantum gates, wherein the circuit depth or circuit width grows rapidly with the size of the data. On the other hand other approaches such as  qGAN have low circuit depth or width, however they are only approximate encoding and suffer from information encoding accuracy. 
    \item The matrices involved in some power system applications (such as Jacobian matrix in power flow or state estimation) may be ill-conditioned. The quantum algorithms require large gate-depths and circuit size to process matrices with poor condition number. To overcome this, preconditioning of matrices or heavy quantum resources may be required. 
    \item Quantum algorithms for power system applications is a  multi-disciplinary topic. Lack of personnel with this specialized skill-set makes improving and implementing reliable quantum algorithms challenging for critical power system  applications. 
\end{itemize}
Further research directions can include (but not limited to) addressing the above challenges for enabling full quantum potential exploitation.
Table \ref{lit_rev_02} provides overview of reviewed power system applications, quantum approaches and potential quantum readiness, giving an indication of whether the corresponding algorithm is NISQ friendly or not.

\begin{landscape} 
\begin{table*}[t]
\centering
\caption{Quantum computing for grid applications}
\label{lit_rev_02}

\begin{tabular}{l l c c l }

\toprule

 \multicolumn{1}{l}{Power System Application} & \multicolumn{1}{c}{References} &\multicolumn{1}{c}{Target} &\multicolumn{1}{c}{Quantum} & \multicolumn{1}{c}{Potential}  \\

 \multicolumn{1}{l}{Sub-Category} &  \multicolumn{1}{c}{} & \multicolumn{1}{c}{Problem} & \multicolumn{1}{c}{Solution} & \multicolumn{1}{c}{{Q readiness}} \\

\cline{1-5}
Static States Estimation & \cite{QC_SA_02} & solve LSE in WLS Gauss-Newton iterations & HHL & No \\
Dynamic States Estimation & \cite{QC_SA_04}\textsuperscript{*}, \cite{QC_SA_05}\textsuperscript{*} & kalman filters & QSDE+QEKF/QKF & No \\
 Observability Analysis &  \cite{QC_SA_01}\textsuperscript{*}  & - & - & - \\
 Meter Placement & \cite{QC_SA_03} & combinatorial optimization & QA & Yes \\
 Power flow &  \cite{QC_SeA_04}, \cite{QC_SeA_05}, \cite{QC_SeA_15} & solve LSE in power flow problem & HHL & No\\

\cline{1-5}
 Contingency Analysis &  \cite{QC_SeA_02} & solve LSE in power flow problem & HHL & No \\
 Cyber Security &  \cite{QC_SeA_06}, \cite{QC_SeA_08}, \cite{QC_SeA_09}, \\
  & \cite{QC_SeA_10}, \cite{QC_SeA_11}, \cite{QC_SeA_12}, \cite{QC_SeA_13}, \cite{QC_SeA_14}  & secure communication & QKD, PQC, QDC & -\\
 Reliability Analysis  & \cite{QC_SeA_01} & Monte-carlo simulation & QAE + QMCS & No\\
\cline{1-5}
Unit commitment & \cite{QC_O_01}, \cite{QC_O_02}, \cite{QC_O_03}, \cite{QC_O_04}, \cite{QC_O_05} & optimization & QAOA, QA & Yes\\
Economic power dispatch & \cite{QC_O_06} & optimization & QPSO & No\\
Facility location allocation  & \cite{QC_O_01} & optimization & VQE, QA & Yes\\
Volt/Var control & \cite{QC_O_07}, \cite{QC_DC_04} & optimization & QGA & No\\
Grid partitioning & \cite{QC_O_08} & optimization & QA & Yes\\
\cline{1-5}
Weather forecast  & \cite{QC_F_01} & - & QSVA & No\\
Generation forecast  & \cite{QC_F_02}, \cite{QC_F_03} & training NN & QGNN, QGA & Yes\\
Load forecast & \cite{QC_F_04}, \cite{QC_F_05}, \cite{QC_F_06}, \cite{QC_F_07} & optimize SVR parameter & QPSO, QBAT, QDA & No\\
Electricity price forecast &  \cite{QC_F_08} & optimize BPNN search space & QIOA & No\\
Predictive maintenance & \cite{QC_F_09} & processing data for NN & HAE & Yes\\
\cline{1-5}
 Small signal Analysis & \cite{QC_St_02}\textsuperscript{*} & Eigenvalue solver & QLS-HHL, QPE, QSVE & No\\
Transient Analysis & \cite{QC_St_01} & data-driven transient stability prediction & VQC as QNN & Yes\\
 EMTP &  \cite{QC_Sim_01}, \cite{QC_Sim_02} & Solve LSE to solve for nodal voltages &  VQLS, HHL & Yes,No\\
 Fault Management &  \cite{QC_FM_01} & fault diagnosis & QC-based deep learning & Yes \\
\cline{1-5}
Load frequency control &  \cite{QC_DC_03} & optimize controller gain values & QIGA, QIEA & Yes\\
DFIG rotor control &  \cite{QC_DC_05} &  optimize rotor side PI controller parameters & QPMMCOA & Yes\\
                   &   \cite{QC_DC_06} & online control strategy updating & QDRL & Yes\\
\bottomrule
\end{tabular}
   \begin{tabular}{ll}   
        \textsuperscript{*}  & only suggestive reference suitable for the corresponding application\\
        \textsuperscript{*}  & {By Q readiness we mean the corresponding algorithm is NISQ friendly or not.}\\
        
  \end{tabular}
\end{table*}
\end{landscape}

\section{Quantum Computation Fundamentals}
Quantum computation is an interdisciplinary branch of physics and computer science which approaches different computational problems using quantum-mechanics, for e.g. entanglement and superposition \cite{Nielsen}. Quantum information and quantum computing  is a center of attention in last few decades due to its ability to outperform classical computation and information processing, because the quantum algorithms can provide significant speedups over their classical counterparts. Many known quantum algorithms have various application, such as:  search algorithm \cite{Grover}, integer factorization \cite{Shor}, solving constraint satisfaction problems \cite{Montanaro}, and quantum machine learning \cite{Mohseni, Biamonte, qsvm}. In this section, a brief summary about basic terminologies and concepts used in quantum computation are outlined.

\subsection{Quantum Hardware and Software}
\subsubsection{Basic unit of information}
The fundamental difference between classical computer and quantum computer is the basic unit of data used for processing of information. In classical computer, the basic unit of data is represented by a bit, which can deterministically assume only one of two possible values/states, that is 0 or 1. Classical computers work by converting information to a series of these bits and process data by manipulating these bits. The classical computers make use of logic gates to perform any operation. 

In quantum computers, the basic unit of data is given by a qubit, which can assume 0 or 1 or any state in between in a probabilistic manner. Currently there exist two main types of quantum computer, quantum annealer and universal quantum computer. 

\subsubsection{Types of quantum computers}
\underline{\textit{Quantum Annealer}} The quantum annealer frames the computational task as an energy minimization problem. Quantum annealers are used for optimization problems; an example is D-Wave quantum annealer.
{Ising machines \cite{Mohseni2022} are hardware solvers (a more generic version of the quantum annealer) which tries to find an absolute or approximate ground state of the Ising model. The Ising machines use three main types of computing methods, namely, classical annealing, quantum annealing and dynamical system evolution. The main limitation of the Ising machine is it's performance is problem-dependent.}

\underline{\textit{Universal Quantum Computer}} {This category of quantum computers implement and compute algorithms with universal quantum gates, analogously to the use of Boolean gates in classical computers. In classical computation the OR, AND, and NOT are the three basic logic gates are called universal gates as they together can construct the logic circuit for any given Boolean expression. Similarly a set of universal quantum gates is any set of gates to which any operation possible on a quantum computer can be reduced, that is, any other unitary operation can be expressed as a finite sequence of gates from the set. For example, Hadamard (H), CNOT, and Phase gates form a set of universal set of quantum gates.} This can be used as general purpose; an example is IBM Quantum machines. 

{The main difference between universal quantum computer and quantum annealer is its applicability. Quantum annealers can be used for a specific purpose of finding an optima, therefore its sole purpose is optimization. On the other hand the universal quantum computer can be used for universal purpose like a regular computer.}

\subsubsection{Quantum simulators} 
{Since the real quantum hardware consume too much resources for prototyping purpose high performance quantum simulators are used. The corresponding software is called Qiskit which is a like a python wrapper for simulating quantum algorithms. In the following few of the examples from IBM quantum simulators are presented. \\
\textbf{AerSimulator and QasmSimulator:} With these simulators we can automatically  mimic an IBM Quantum backend (real hardware). With this simulator we can configure the same basis gates, coupling map, and basic noise model for that specific backend.\\
\textbf{StatevectorSimulator and UnitarySimulator:} This simulators operate computing actual state vectors (for quantum states) and unitary matrices (for operators) for an ideal quantum circuit simulation using local CPU or GPU.}

\subsection{Quantum state, qubit, and it's representation}

A quantum state is any possible state in which a quantum mechanical system can be. A fully specified quantum state can be described by a state vector, a wavefunction, or a complete set of quantum numbers for a specific system. {A partially known quantum state can be described by a density matrix (or density operator) \cite{Nielsen}}.

\begin{figure}[!ht]
    \centering
    \includegraphics[width=0.2\textwidth]{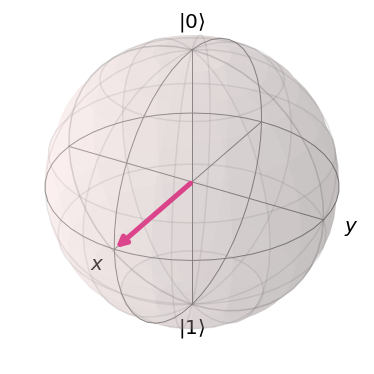}
    \caption[Caption used in list of tables]{Representation of the $|{+}\rangle$ state in Bloch sphere.}
    \label{fig:bloch_sphere}
\end{figure}

There exists a visual representation of a qubit state called Bloch sphere representation. Suppose we want to plot our general one qubit state:
\begin{equation}
| \alpha \rangle = \cos{\tfrac{\theta}{2}}|0\rangle + e^{i\phi}\sin{\tfrac{\theta}{2}}|1\rangle. \label{qubit_state} 
\end{equation}
If we interpret $\theta$ and $\phi$ as spherical co-ordinates ($r = 1$, since the magnitude of the qubit state is $1$), we can plot any single qubit state on the surface of a sphere, known as the Bloch sphere. In Fig. \ref{fig:bloch_sphere}, we have plotted a qubit in the state $|{+}\rangle =\tfrac{1}{\sqrt{2}}|0\rangle + \tfrac{1}{\sqrt{2}}|1\rangle$. In this case, $\theta = \pi/2$ and $\phi = 0$.

\subsection{System of qubits}

A joint state of multiple qubits can be described using tensor product. 
For e.g. from two one qubit states $\scriptstyle \ket{\alpha} = \scriptstyle \begin{pmatrix}\alpha_0\\\alpha_1\end{pmatrix}$ and $\scriptstyle \ket{\beta} = \scriptstyle \begin{pmatrix}\beta_0\\\beta_1\end{pmatrix}$ we can build a two qubit state:
\begin{equation}
\ket{\alpha\beta} =\ket{\alpha}\otimes\ket{\beta}  = \alpha_0\beta_0\ket{00} + \alpha_0\beta_1\ket{01} + \alpha_1\beta_0\ket{10} + \alpha_1\beta_1\ket{11}.
    \label{two_qubit_stat}
\end{equation}

However a generic two-qubit state can be written as \begin{equation}
\ket{\gamma}= \gamma_{00}\ket{00} + \gamma_{01}\ket{01} + \gamma_{10}\ket{10} + \gamma_{11}\ket{11}. \label{superposition}
\end{equation}
For a system of $n$ qubits, the number of basis vectors is  $2^n$. Therefore, the dimension of the state space grows exponentially ($2^n$) with the number of qubits $n$. 

The most important properties of Qubits are the superposition and entanglement of states, allowing the execution of exponentially many computations in parallel.

\subsubsection{Superposition}

In quantum mechanics superposition is the property of a quantum system to be in a multiple states at the same time.  This means that with $n$ qubits, one can operate on $2^n$ numbers simultaneously.

Translation to Quantum Computation: With regards to qubits manipulation, superposition refers to the fact that any linear combination of two quantum states, once normalized, will also be a valid quantum state. In other words, any quantum state can be expressed as a linear combination of a few basis states \cite{QC_Beginners_guide}. For e.g., in Eq.(\ref{superposition}), $\ket{\gamma}$ represents a generic two qubit state with a linear superposition of the four two-qubit basis states.

\subsubsection{Entanglement}

In quantum mechanics, for certain pairs of particles, irrespective of the position of particles in space, the state of one particle is complementary to the state of the other, and therefore have quantum states that are linked together. The coupling of atomic particles in this way is known as entanglement.

Translation to quantum computation: States of a system which cannot be expressed as a tensor product of states of its individual subsystems are called entangled states. For a system of $n$ qubits, this means that an entalged state cannot be written a tensor product of $n$ single qubit states. For instance, for system of two qubits, the state $\frac{\ket{00}+\ket{11}}{\sqrt{2}}$ is an entangled state, as this state cannot be written as tensor product of two single qubit state \cite{QC_guide}.
   
{\subsection{Observable and measurement}
Beside qubits, the two important concepts of quantum mechanics are observable and measurement which is essential for quantum computing.\\
\textbf{Observable:}
In quantum mechanics, an observable is a physical quantity that can be measured, e.g., position, momentum, or energy. In classical mechanics, it is a real-valued function on the set of all possible system states. In quantum physics, it is an operator, usually represented by a unitary matrix, where the property of the quantum state can be determined by some sequence of operations, e.g., energy of a quantum system is represented by Hamiltonian operator.\\
\textbf{Measurement:}\\
Measurement is a non unitary operation after which a generic quantum state collapsed into one of its basis states; therefore measurement outcomes are probabilistic.  
For example, after a measurement, the one-qubit state $\ket{\alpha}$ in \eqref{qubit_state} will collapse into either $\ket{0}$-state with a probability of $|\alpha_0|^2$ or $\ket{1}$-state with a probability of $|\alpha_1|^2$, with a condition $|\alpha_0|^2 + |\alpha_1|^2 = 1 $. 
}

\subsection{Steps in building quantum algorithm}
There are three main steps for building a generic quantum algorithm.   
\begin{itemize}
    \item \textit{Quantum information encoding :}  The first step is loading the classical data in a quantum computer. While efficient data encoding is still an open problem but few popular examples for data encoding are: \\
    {(1) The first one is the \textit{qRAM}, namely a quantum version of the random access memory (RAM) \cite{Lloyd}. From a theoretical perspective, it can be described as a quantum operator allowing efficient (i.e., quantum-parallel) access to classically stored information. Unfortunately though such a device was never built at scale. \\
    (2) The second approach is preparing the desired quantum state directly with controlled rotation gates \cite{ barenco, Kumar, Cortes, Plesch, ventura1999initializing, araujo2021divide, araujo_configurable_2022}. The main challenges for an efficient encoding in this method is optimizing the  trade off between the  circuit-depth and circuit-width.\\ 
    (3) The third approach resorts to \textit{qGAN}, a quantum version of Generative adversarial network (GAN) \cite{zoufal2019quantum, herr_anomaly_2020, chakrabarti_quantum_2019} etc. Out of different encoding, qGAN encoding stands out due to its qubit-efficiency and its capacity to be prepared in a polynomial number of gates using qGANs. {This is a popular choice where parametrised quantum circuit with parameter $\theta$ to generate the $n$-qubit quantum state  \begin{equation}
    \ket{ \psi (\theta) } _n = \sum_{i=0}^{2^n-1} \sqrt{p_i (\theta)} \ket{ i }_n,
    \end{equation}
    where $\ket{ i }_n$ are the basis vectors and $\sqrt{p_i (\theta)}$ are the corresponding amplitude.}
    In a qGAN the classical generator or discriminator or both can be a parametric quantum circuit.}
    
    \item \textit{Quantum information processing:} After the data loading the second step is to implement the actual algorithm, for e.g.,  Grover \cite{Grover} or Shor \cite{Shor} algorithm. 
    
    \item \textit{Quantum information decoding:} After the quantum computation the third step is to extract the useful information from the quantum computer. We get a normalized vector (or quantum state) after the quantum simulation, therefore, we have to post processing for getting the actual unscaled result.

\end{itemize}

Readers curious to gain more fundamental knowledge and hands-on experience on quantum computation may refer to IBM's extensive documentation and learning platform on \cite{ibm_quantum_doc} and \cite{ibm_quantum_learning}. Quantum algorithms can be implemented by building a circuit in IBM composer \cite{ibm_quantum_composer} or by writing lines of code using the Qiskit package \cite{qiskit2024}. 

\section{Conclusion}
In recent years, the use of quantum  quantum algorithms for power system applications have gained traction due to both increasing complexity and computational demand of the applications as well as advancements in quantum technology. However the research in this direction is still in nascent stages. This paper serves as a primer for the reader by emphasizing the importance of research on quantum solutions for power system application and providing a comprehensive review on existing state-of-the art quantum algorithms for various power system applications.  
Following this the paper highlights the challenges in exploiting full quantum potential and provides an indication on  quantum readiness level of the reviewed power system applications. Furthermore, applications are identified which can benefit from quantum computation, however quantum algorithms are yet to be implemented. Finally, quantum fundamentals are introduced to serve as a stepping stone for further research in this area.



\bibliographystyle{IEEEtran}
\bibliography{IEEEabrv,access}

\begin{thebibliography}{100}
\providecommand{\url}[1]{#1}
\csname url@samestyle\endcsname
\providecommand{\newblock}{\relax}
\providecommand{\bibinfo}[2]{#2}
\providecommand{\BIBentrySTDinterwordspacing}{\spaceskip=0pt\relax}
\providecommand{\BIBentryALTinterwordstretchfactor}{4}
\providecommand{\BIBentryALTinterwordspacing}{\spaceskip=\fontdimen2\font plus
\BIBentryALTinterwordstretchfactor\fontdimen3\font minus \fontdimen4\font\relax}
\providecommand{\BIBforeignlanguage}[2]{{%
\expandafter\ifx\csname l@#1\endcsname\relax
\typeout{** WARNING: IEEEtran.bst: No hyphenation pattern has been}%
\typeout{** loaded for the language `#1'. Using the pattern for}%
\typeout{** the default language instead.}%
\else
\language=\csname l@#1\endcsname
\fi
#2}}
\providecommand{\BIBdecl}{\relax}
\BIBdecl

\bibitem{QC_PA_01}
R.~Eskandarpour, A.~Khodaei, L.~Zhang, E.~A. Paaso, and S.~Bahramirad, ``Quantum computing applications in power systems,'' \emph{CIGRE US National Committee}, vol. 2019 Grid of the Future Symposium, 2019.

\bibitem{QC_PA_10}
S.~Golestan, M.~R. Habibi, S.~Y.~M. Mousavi, J.~M. Guerrero, and J.~C. Vasquez, ``Quantum computation in power systems: An overview of recent advances,'' \emph{Energy Reports}, vol.~9, pp. 584--596, 2023.

\bibitem{QC_PA_06}
H.~P. Paudel, M.~Syamlal, S.~E. Crawford, Y.-L. Lee, R.~A. Shugayev, P.~Lu, P.~R. Ohodnicki, D.~Mollot, and Y.~Duan, ``Quantum computing and simulations for energy applications: Review and perspective,'' \emph{ACS Engineering Au}, vol.~2, no.~3, pp. 151--196, 2022.

\bibitem{RES_news_01}
Eurostat, ``Renewable energy statistics,'' \emph{eurostat: statistics explained}, 2021.

\bibitem{QC_news_03}
\BIBentryALTinterwordspacing
V.~Brownell. Quantum computing could change the way the world uses energy. Available: \underline. [Online]. Available: \url{https://qz.com/1566061/quantum-computing-will-change-the-way-the-world-uses-energy}
\BIBentrySTDinterwordspacing

\bibitem{QC_news_02}
Ibm, ``Our new 2022 development roadmap,'' \emph{IBM}, 2022.

\bibitem{QC_news_04}
\BIBentryALTinterwordspacing
J.~Hsu. How much power will quantum computing need? Available: \underline. [Online]. Available: \url{https://spectrum.ieee.org/how-much-power-will-quantum-computing-need}
\BIBentrySTDinterwordspacing

\bibitem{QC_news_01}
Epri, ``Quantum computing: Technology update across the energy industry,'' \emph{EPRI}, 2020.

\bibitem{QC_SA_02}
F.~Feng, P.~Zhang, Y.~F. Zhou, and Z.~F. Tang, ``Quantum microgrid state estimation,'' \emph{Electric Power Systems Research}, vol. 212, p. 108386, 2022.

\bibitem{C_SA_02}
J.~Zhao, A.~G. Exposito, M.~Netto, L.~Mili, A.~Abur, V.~Terzija, I.~Kamwa, B.~Pal, A.~K. Singh, Z.~Huang, and A.~P.~S. Meliopoulos, ``Power system dynamic state estimation: Motivations, definitions, methodologies, and future work,'' \emph{IEEE Transactions on Power Systems}, vol.~34, pp. 3188--3198, 2019.

\bibitem{taskforce4}
J.~Zhao, A.~K. Singh, A.~S. Mir, A.~Taha, A.~Rouhani, A.~Gomez-Exposito, A.~Meliopoulos, B.~Pal, I.~Kamwa, J.~Qi, L.~Mili, M.~A. Mohd~Ariff, M.~Netto, M.~Glavic, S.~Yu, S.~Wang, T.~Bi, T.~Van~Cutsem, V.~Terzija, and Z.~Huang, ``Power system dynamic state and parameter estimation-transition to power electronics-dominated clean energy systems,'' \emph{IEEE PES Technical Report, TR88}, June 2021.

\bibitem{C_SA_03}
Y.~Liu, A.~K. Singh, J.~Zhao, A.~S.~P. Meliopoulos, B.~Pal, M.~A. B.~M. Ariff, T.~V. Custem, M.~Glavic, Z.~Huang, I.~Kamwa, L.~Mili, A.~S. Mir, A.~Taha, V.~Terzija, and S.~Yu, ``Dynamic state estimation for power system control and protection,'' \emph{IEEE Transactions on Power Systems}, vol.~36, no.~6, pp. 5909--5921, 2021.

\bibitem{QC_SA_05}
L.~Accardi, ``Quantum kalman filters,'' in \emph{Mathematical System Theory: The Influence of R. E. Kalman, Berlin, : Berlin Heidelberg}.\hskip 1em plus 0.5em minus 0.4em\relax Springer, 1991, pp. 135--143.

\bibitem{QC_SA_04}
M.~F. Emzir, M.~J. Woolley, and I.~R. Petersen, ``A quantum extended kalman filter,'' \emph{arXiv [quant-ph]}, 2016.

\bibitem{Abur_Exposito}
A.~Abur and A.~G. Exposito, \emph{Power System State Estimation Theory and Implementation}.\hskip 1em plus 0.5em minus 0.4em\relax Inc: Marcel Dekker, 2004.

\bibitem{QC_SA_01}
Y.~Tang, J.~Yan, and H.~Edwin, ``From quantum graph computing to quantum graph learning: A survey,'' \emph{arXiv [quant-ph]}, 2022.

\bibitem{C_SA_01}
M.~Zeraati, M.~Shabanzadeh, M.~R. Sheibania, and F.~Jabari, ``Meter placement algorithms to enhance distribution systems state estimation: Review, challenges and future research directions,'' \emph{IET Renewable Power Generation}, 2022.

\bibitem{QC_SA_03}
E.~B. Jones, E.~Kapit, C.~Y. Chang, D.~Biagioni, D.~Vaidhynathan, P.~Graf, and W.~Jones, ``On the computational viability of quantum optimization for pmu placement,'' in \emph{2020 IEEE Power and Energy Society General Meeting (PESGM)}, 2020.

\bibitem{C_SeA_01}
Epri, \emph{Contingency Analysis - Baseline}.\hskip 1em plus 0.5em minus 0.4em\relax IECSA Volume II, 2010.

\bibitem{QC_SeA_04}
R.~Eskandarpour, K.~Ghosh, A.~Khodaei, L.~Zhang, A.~Paaso, and S.~Bahramirad, ``Quantum computing solution of dc power flow,'' \emph{arXiv [quant-ph]}, 2020.

\bibitem{QC_SeA_04_1}
R.~Eskandarpour, K.~Ghosh, A.~Khodaei, and A.~Paaso, ``Experimental quantum computing to solve network dc power flow problem,'' \emph{arXiv[quant-ph]}, 2021.

\bibitem{QC_SeA_05}
F.~Fend, Y.~Zhou, and P.~Zhang, ``Quantum power flow,'' \emph{IEEE Transactions on Power Systems}, vol.~36, pp. 3810--3812, 2021.

\bibitem{QC_SeA_02}
R.~Eskandarpour, P.~Gokhale, A.~Khodaei, F.~T. Chong, and S.~B. E.~Passo, ``Quantum computing for enhancing grid security,'' \emph{IEEE Transactions on Power Systems}, vol.~35, no.~4, pp. 4135--4137, 2022.

\bibitem{QC_SeA_15}
B.~Sævarsson, S.~Chatzivasileiadis, H.~Jóhannsson, and J.~Østergaard, ``Quantum computing for power flow algorithms: Testing on real quantum computers,'' \emph{arXiv [quant-ph]}, 2022.

\bibitem{QC_SeA_07}
P.~W. Shor, ``Algorithms for quantum computation: Discrete logarithms and factoring,'' in \emph{Proceedings of the 35th Annual Symposium on Foundations of Computer Science, 1994}, 1994.

\bibitem{QC_SeA_06}
J.~Ahn, J.~Chung, T.~Kim, B.~Ahn, and J.~Choi, ``An overview of quantum security for distributed energy resources,'' in \emph{2021 IEEE 12th International Symposium on Power Electronics for Distributed Generation Systems (PEDG)}, 2021.

\bibitem{QC_SeA_14}
M.~Alshowkan, P.~G. Evans, M.~Starke, D.~Earl, and N.~A. Peters, ``Authentication of smart grid communications using quantum key distribution,'' \emph{Scientific Reports}, vol.~12, 2022.

\bibitem{QC_SeA_08}
Z.~Tang, Y.~Qin, Z.~Jiang, W.~O. Krawec, and P.~Zhang, ``Quantum-secure microgrid,'' \emph{IEEE Transactions on Power Systems}, vol.~36, pp. 1250--1263, 2021.

\bibitem{QC_SeA_09}
{Z. Tang and Y. Qin and Z. Jiang and W. O. Krawec and P. Zhang}, ``Quantum-secure networked microgrid,'' in \emph{2020 IEEE Power and Energy Society General Meeting (PESGM)}, 2020.

\bibitem{QC_SeA_10}
Z.~Tang, P.~Zhang, W.~O. Krawec, and Z.~Jiang, ``Programmable quantum networked microgrids,'' \emph{IEEE Transactions on Quantum Engineering}, vol.~1, 2020.

\bibitem{QC_SeA_11}
Z.~Tang, P.~Zhang, and W.~O. Krawec, ``A quantum leap in microgrids security: The prospects of quantum-secure microgrids,'' \emph{IEEE Electrification Magazine}, vol.~9, 2021.

\bibitem{QC_PA_02}
Y.~Zhou, Z.~Tang, N.~Nikmehr, P.~Babahajiyani, F.~Feng, T.~C. Wei, H.~Zheng, and P.~Zhang, ``Quantum computing in power systems,'' \emph{iEnergy}, vol.~1, 2022.

\bibitem{QC_SeA_12}
Z.~Tang, P.~Zhang, W.~O. Krawec, and L.~Wang, ``Quantum networks for resilient power grids: Theory and simulated evaluation,'' \emph{IEEE Transactions on Power Systems (Early Access)}, 2022.

\bibitem{QC_SeA_13}
Z.~M. Jiang, Z.~Tang, Y.~Qin, C.~Kang, and P.~Zhang, ``Quantum internet for resilient electric grids,'' \emph{International Transactions on Electrical Energy Systems}, vol.~31, 2021.

\bibitem{QC_DC_01}
P.~Babahajiani and P.~Zhang, ``Quantum distributed microgrid control,'' in \emph{2022 IEEE Power and Energy Society General Meeting (PESGM)}, 2022.

\bibitem{QC_DC_02}
{P. Babahajiani and P. Zhang}, ``Quantum-secure distributed frequency control,'' in \emph{2022 IEEE Power and Energy Society General Meeting (PESGM)}, 2022.

\bibitem{QC_SeA_01}
N.~Nikmehr and P.~Zhang, ``Quantum distribution system reliability assessment,'' in \emph{2022 IEEE Power and Energy Society General Meeting (PESGM)}, 2022.

\bibitem{QC_SeA_03}
S.~You, ``A quantum computing framework for complex system reliability assessment,'' \emph{arXiv [eess.sy]}, 2021.

\bibitem{QC_O_01}
A.~Ajagekar and F.~You, ``Quantum computing for energy systems optimization: Challenges and opportunities,'' \emph{Energy}, vol. 179, pp. 76--89, 2019.

\bibitem{QC_O_02}
N.~Nikmehr, P.~Zhang, and M.~A. Bragin, ``Quantum distributed unit commitment: An application in microgrids,'' \emph{IEEE Transactions on Power Systems}, vol.~37, pp. 3592--3603, 2022.

\bibitem{QC_O_03}
{N. Nikmehr and P. Zhang and M. A. Bragin}, ``Quantum-enabled distributed unit commitment,'' in \emph{2022 IEEE Power and Energy Society General Meeting (PESGM)}, 2022.

\bibitem{QC_O_04}
S.~Koretsky \emph{et~al.}, ``Adapting quantum approximation optimization algorithm (qaoa) for unit commitment,'' in \emph{2021 IEEE International Conference on Quantum Computing and Engineering (QCE)}, 2021.

\bibitem{QC_O_05}
F.~Feng, P.~Zhang, M.~A. Bragin, and Y.~Zhou, ``Novel resolution of unit commitment problems through quantum surrogate lagrangian relaxation,'' \emph{IEEE Transactions on Power Systems (Early Access)}, 2022.

\bibitem{BENDERS}
J.~F. Benders, ``Partitioning procedures for solving mixed-variables programming problems,'' \emph{Numerische Mathematik}, vol.~4, pp. 238--252, 1963.

\bibitem{Rahmaniani}
R.~Rahmaniani, T.~G. Crainic, M.~Gendreau, and W.~Rei, ``The benders decomposition algorithm: A literature review,'' \emph{European Journal of Operational Research}, vol. 259, no.~3, pp. 801--817, 2017.

\bibitem{Zhao}
Z.~Zhao, L.~Fan, and Z.~Han, ``Hybrid quantum benders,'' in \emph{Decomposition For Mixed-integer Linear Programming', 2022 IEEE Wireless Communications and Networking Conference (WCNC)}, 2021, pp. 2536--2540.

\bibitem{Gao}
F.~Gao, D.~Huang, Z.~Zhao, W.~Dai, M.~Yang, and F.~Shuang, ``Hybrid quantum-classical general benders decomposition algorithm for unit commitment with multiple networked microgrids,'' 2022, [quant-ph].

\bibitem{Nikolaidis}
P.~Nikolaidis and S.~Chatzis, ``{G}aussian process-based {B}ayesian optimization for data-driven unit commitment,'' \emph{International Journal of Electrical Power and Energy Systems}, vol. 130, 2024.

\bibitem{Rapp2023}
F.~Rapp and M.~Roth, ``Quantum {G}aussian process regression for {B}ayesian optimization,'' \emph{Quantum Mach. Intell.}, vol.~6, p.~5, 2024.

\bibitem{Dai2024}
Z.~Dai, G.~K.~R. Lau, A.~Verma, Y.~Shu, B.~K.~H. Low, and P.~Jaillet, ``Quantum bayesian optimization. advances in neural information processing systems,'' \emph{36}, 2024.

\bibitem{QC_O_06}
L.~Xu, L.~Zhang, and J.~Dang, ``Quantum-behaved particle swarm optimization with cauchy disturbance for power economic dispatch,'' in \emph{2012 4th International Conference on Intelligent Human-Machine Systems and Cybernetics}, 2012.

\bibitem{QC_O_07}
H.~Miao, H.~Wang, and Z.~Deng, ``Quantum genetic algorithm and its application in power system reactive power optimization,'' in \emph{2009 International Conference on Computational Intelligence and Security}, 2009.

\bibitem{QC_DC_04}
J.~G. Vlachogiannis and J.~Østergaard, ``Reactive power and voltage control based on general quantum genetic algorithms,'' \emph{Expert Systems with Applications}, vol.~36, no.~3, pp. 6118--6126, 2009.

\bibitem{QC_O_08}
M.~Fernández-Campoamor, C.~O'Meara, G.~Cortiana, V.~Peric, and J.~Bernabé-Moreno, ``Community detection in electrical grids using quantum annealing,'' \emph{arXiv [quant-ph]}, 2021.

\bibitem{QC_F_01}
M.~Senekane and B.~M. Taele, ``Prediction of solar irradiation using quantum support vector machine learning algorithm,'' \emph{Smart Grid, Renew. Energy}, vol.~7, 2016.

\bibitem{QC_F_02}
D.~K. Chaturvedi and A.~Yadav, ``Forecasting of solar power using quantum ga - gnn,'' \emph{International Journal of Computer Applications}, vol. 128, 2015.

\bibitem{QC_F_03}
Z.~Kou, T.~Liu, and J.~Zhao, ``Generation prediction of ultra-short-term wind farm based on quantum genetic algorithm and fuzzy neural network,'' in \emph{2020 39th Chinese Control Conference (CCC)}, 2020.

\bibitem{QC_F_04}
Z.~Zhang and W.~C. Hong, ``Electric load forecasting by complete ensemble empirical mode decomposition adaptive noise and support vector regression with quantum-based dragonfly algorithm,'' \emph{Nonlinear Dyn}, vol.~98, pp. 1107--1136, 2019.

\bibitem{QC_F_05}
M.~L. Huang, ``Hybridization of chaotic quantum particle swarm optimization with svr in electric demand forecasting,'' \emph{Energies}, vol.~9, 2016.

\bibitem{QC_F_06}
L.~L. Peng, G.~F. Fan, M.~L. Huang, and W.~C. Hong, ``Hybridizing demd and quantum pso with svr in electric load forecasting,'' \emph{Energies}, vol.~9, 2016.

\bibitem{QC_F_07}
M.~W. Li, J.~Geng, S.~Wang, and W.~C. Hong, ``Hybrid chaotic quantum bat algorithm with svr in electric load forecasting,'' \emph{Energies}, vol.~10, 2017.

\bibitem{QC_F_08}
X.~Zhang, Q.~Hao, W.~Qu, X.~Ji, Y.~Zhang, and B.~Xu, ``Electricity price forecasting method based on quantum immune optimization bp neural network algorithm,'' in \emph{2021 6th Asia Conference on Power and Electrical Engineering (ACPEE)}, 2021.

\bibitem{QC_F_09}
A.~Sakhnenko, C.~O'Meara, K.~J.~B. Ghosh, C.~B. Mendl, G.~Cortiana, and J.~Bernabé-Moreno, ``Hybrid classical-quantum autoencoder for anomaly detection,'' \emph{Quantum Mach. Intell.}, vol.~4, p.~27, 2021.

\bibitem{C_St_03}
P.~Kundur, \emph{Power System Stability and Control}.\hskip 1em plus 0.5em minus 0.4em\relax Inc: McGraw-Hill, 1993.

\bibitem{C_St_02}
P.~W. Sauer and M.~A. Pai, \emph{Power System Dynamics and Stability}.\hskip 1em plus 0.5em minus 0.4em\relax Prentice Hall, 2006.

\bibitem{QC_St_02}
C.~Shao, ``Computing eigenvalues of matrices in a quantum computer,'' \emph{arXiv [quant-ph]}, 2019.

\bibitem{C_Sim_01}
H.~W. Dommel, \emph{EMTP Theory Book}.\hskip 1em plus 0.5em minus 0.4em\relax Vancouver, BC, Canada: Microtran Power System Analysis Corporation, 1996.

\bibitem{QC_Sim_01}
Y.~Zhou, F.~Feng, and P.~Zhang, ``Quantum electromagnetic transients program,'' \emph{IEEE Transactions on Power Systems}, vol.~36, pp. 3813--3816, 2021.

\bibitem{QC_Sim_02}
Y.~Zhou, P.~Zhang, and F.~Feng, ``Noisy-intermediate-scale quantum electromagnetic transients program,'' \emph{IEEE Transactions on Power Systems (Early Access)}, 2022.

\bibitem{C_St_01}
H.~D. Chiang and L.~F. Alberto, \emph{Stability Regions of Nonlinear Dynamical Systems: Theory, Estimation, and Applications}.\hskip 1em plus 0.5em minus 0.4em\relax Cambridge University Press, 2015.

\bibitem{QC_St_01}
Y.~Zhou and P.~Zhang, ``Noise-resilient quantum machine learning for stability assessment of power systems,'' \emph{arXiv [quant-ph]}, 2021.

\bibitem{QC_FM_01}
A.~Ajagekar and F.~You, ``Quantum computing based hybrid deep learning for fault diagnosis in electrical power systems,'' \emph{Applied Energy}, vol. 303, p. 117628, 2021.

\bibitem{QC_DC_03}
K.~Jagatheesan, S.~Samanta, A.~Choudhury, N.~Dey, B.~Anand, and A.~S. Ashour, ``Quantum inspired evolutionary algorithm in load frequency control of multi-area interconnected thermal power system with non-linearity,'' in \emph{2022 IEEE Power and Energy Society General Meeting (PESGM)}, 2022.

\bibitem{QC_DC_05}
K.~Han, T.~Huang, and L.~Yin, ``Quantum parallel multi-layer monte carlo optimization algorithm for controller parameters optimization of doubly-fed induction generator-based wind turbines,'' \emph{Applied Soft Computing}, vol. 112, p. 107813, 2021.

\bibitem{QC_DC_06}
L.~Yin, L.~Chen, D.~Liu, X.~Huang, and F.~Gao, ``Quantum deep reinforcement learning for rotor side converter control of double-fed induction generator-based wind turbines,'' \emph{Engineering Applications of Artificial Intelligence}, vol. 106, p. 104451, 2021.

\bibitem{Shor}
P.~W. Shor, ``Algorithms for quantum computation: discrete logarithms and factoring,'' in \emph{Proceedings 35th annual symposium on foundations of computer science}, 1994, pp. 124--134.

\bibitem{Grover}
L.~K. Grover, ``Quantum mechanics helps in searching for a needle in a haystack,'' \emph{Phys. Rev. Lett.}, vol.~79, no.~2, p. 325, 1997.

\bibitem{Nielsen}
M.~A. Nielsen and C.~I. L., \emph{Quantum Computation and Quantum Information}.\hskip 1em plus 0.5em minus 0.4em\relax Cambridge: Cambridge University Press, 2001.

\bibitem{Montanaro}
A.~Montanaro, ``Quantum walk speedup of backtracking algorithms,'' 2015, 02374 [quant-ph].

\bibitem{Mohseni}
S.~Lloyd, M.~Mohseni, and P.~Rebentrost, ``Quantum algorithms for supervised and unsupervised machine learning,'' 2013, 0411 [quant-ph].

\bibitem{Biamonte}
J \emph{et~al.}, ``Biamonte, `quantum machine learning','' \emph{Nature}, vol. 549, no. 7671, pp. 195--202, 2017.

\bibitem{qsvm}
P.~Rebentrost, M.~Mohseni, and S.~Lloyd, ``Quantum support vector machine for big data classification,'' \emph{Physical review letters}, vol. 113, no.~13, p. 130503, 2014.

\bibitem{Mohseni2022}
N.~Mohseni, P.~L. McMahon, and T.~Byrnes, ``Ising machines as hardware solvers of combinatorial optimization problems,'' \emph{Nat. Rev. Phys.}, vol.~4, pp. 363--379, 2022.

\bibitem{QC_Beginners_guide}
J.~Abhijith, A.~Adedoyin, J.~Ambrosiano, P.~Anisimov, W.~Casper, G.~Chennupati, C.~Coffrin, H.~Djidjev, D.~Gunter, S.~Karra, N.~Lemons, S.~Lin, A.~Malyzhenkov, D.~Mascarenas, S.~Mniszewski, B.~Nadiga, D.~O'malley, D.~Oyen, S.~Pakin, L.~Prasad, R.~Roberts, P.~Romero, N.~Santhi, N.~Sinitsyn, P.~J. Swart, J.~G. Wendelberger, B.~Yoon, R.~Zamora, W.~Zhu, S.~Eidenbenz, A.~B\"{a}rtschi, P.~J. Coles, M.~Vuffray, and A.~Y. Lokhov, ``Quantum algorithm implementations for beginners,'' \emph{ACM Transactions on Quantum Computing}, vol.~3, no.~4, pp. 1--92, 2022.

\bibitem{QC_guide}
M.~A. Nielsen and I.~L. Chuang, \emph{Quantum Computation and Quantum Information: 10th Anniversary Edition}.\hskip 1em plus 0.5em minus 0.4em\relax Cambridge University Press, 2010.

\bibitem{Lloyd}
V.~Giovannetti, S.~Lloyd, and L.~Maccone, ``Quantum random access memory,'' \emph{Phys. Rev. Lett.}, vol. 100, no.~16, p. 160501, 2008.

\bibitem{barenco}
A.~Barenco \emph{et~al.}, ``Elementary gates for quantum computation,'' \emph{Physical review A}, vol.~52, no.~5, p. 3457, 1995.

\bibitem{Kumar}
P.~Kumar, ``Direct implementation of an n-qubit controlled-unitary gate in a single step,'' \emph{Quantum information processing}, vol.~12, no.~2, pp. 1201--1223, 2013.

\bibitem{Cortes}
J.~A. Cortese and T.~M. Braje, ``Loading classical data into a quantum computer,'' 1958, [quant-ph]. 2018.

\bibitem{Plesch}
M.~Plesch and {\v{C}}.~. Brukner, ``Quantum-state preparation with universal gate decompositions,'' \emph{Phys. Rev. A}, vol.~83, no.~3, p. 032302, 2011.

\bibitem{ventura1999initializing}
D.~Ventura and T.~Martinez, ``Initializing the amplitude distribution of a quantum state,'' \emph{Foundations of Physics Letters}, vol.~12, no.~6, pp. 547--559, 1999.

\bibitem{araujo2021divide}
I.~F. Araujo, D.~K. Park, F.~Petruccione, and A.~J. da~Silva, ``A divide-and-conquer algorithm for quantum state preparation,'' \emph{Scientific Reports}, vol.~11, no.~1, pp. 1--12, 2021.

\bibitem{araujo_configurable_2022}
I.~F. Araujo, D.~K. Park, T.~B. Ludermir, W.~R. Oliveira, F.~Petruccione, and A.~J. da~Silva, ``Configurable sublinear circuits for quantum state preparation,'' 2022, 10182 [quant-ph].

\bibitem{zoufal2019quantum}
C.~Zoufal, A.~Lucchi, and S.~Woerner, ``Quantum generative adversarial networks for learning and loading random distributions,'' \emph{npj Quantum Information}, vol.~5, no.~1, pp. 1--9, 2019.

\bibitem{herr_anomaly_2020}
D.~Herr, B.~Obert, and M.~Rosenkranz, ``Anomaly detection with variational quantum generative adversarial networks,'' \emph{Quantum Science and Technology}, vol.~6, no.~4, p. 045004, 2021.

\bibitem{chakrabarti_quantum_2019}
S.~Chakrabarti, H.~Yiming, T.~Li, S.~Feizi, and X.~Wu, ``Quantum wasserstein generative adversarial networks,'' \emph{Advances in Neural Information Processing Systems}, vol.~32, 2019.

\bibitem{ibm_quantum_doc}
\BIBentryALTinterwordspacing
(2023) {IBM Quantum Documentation}. [Online]. Available: \url{https://docs.quantum.ibm.com}
\BIBentrySTDinterwordspacing

\bibitem{ibm_quantum_learning}
\BIBentryALTinterwordspacing
(2023) {IBM Quantum Learning}. [Online]. Available: \url{https://learning.quantum.ibm.com/}
\BIBentrySTDinterwordspacing

\bibitem{ibm_quantum_composer}
\BIBentryALTinterwordspacing
(2023) {IBM Quantum Composer}. [Online]. Available: \url{https://quantum.ibm.com/composer}
\BIBentrySTDinterwordspacing

\bibitem{qiskit2024}
A.~Javadi-Abhari, M.~Treinish, K.~Krsulich, C.~J. Wood, J.~Lishman, J.~Gacon, S.~Martiel, P.~D. Nation, L.~S. Bishop, A.~W. Cross, B.~R. Johnson, and J.~M. Gambetta, ``Quantum computing with {Q}iskit,'' 2024.

\end{thebibliography}

\end{document}